\documentstyle[epsfig,fleqn,twocolumn,twoside]{article}
\begin{document}

\pagestyle{myheadings}

\markboth{{\it Journal  of Experimental and Theoretical Physics}, Vol.
95,   No.1,    2002,    pp.   5-10}{Translated   from   {\it   Zhurnal
Eksperimentai'noi i Teoreticheskoi Fiziki}, Vol. 122,  No.1, 2002, pp.
10-16}

\twocolumn[\hsize\textwidth\columnwidth\hsize\csname
@twocolumnfalse\endcsname

\title{Twenty Years of  Galactic  Observations in Searching for Bursts
of  Collapse  Neutrinos  with  the  Baksan  Underground  Scintillation
Telescope}

\author{ {E.N.Alexeyev${}^*$, L.N.Alexeyeva}  \\
\it     Institute for Nuclear Research, Russian Academy of Sciences, \\
\it     pr.Shestidesyatiletiya Oktyabrya 7a, Moscow, 117312 Russia   \\
\it               ${}^*$e-mail: alexeyev@ms1.inr.ac.ru               \\
\it               Received: January 29, 2002}

\date{}

\maketitle

\begin{abstract}
The results  of  twenty-year-long  Galactic  observations  in neutrino
radiation  are  summarized.  Except  for the recording of  a  neutrino
signal from the  supernova SN 1987A,  no Galactic bursts  of  collapse
neutrinos have been detected. An upper bound on the mean  frequency of
gravitational collapses  in  our  Galaxy  was  obtained, $f_{collapse}
(90\% \mbox{confidence}) < 0.13 \mbox{yr}^{-1}$.
\end{abstract}


\vskip5mm

]

\section{Introduction}

In 1933, Wolfgang Pauli introduced  a  new neutral particle of low  or
zero mass  to save  the law of conservation of  energy in nuclear beta
decays \cite{c01}.  This  particle  was  experimentally  observed only
twenty--five  years  later   by   Reines  and  Cowan  \cite{c02}.  The
interactions  of  neutrinos  and  their  role   in  particle  physics,
astrophysics,  and  cosmology  have  been studied with  an  increasing
intensity ever since.  The  discovered particle  turned  out to be  so
amazing that it  allowed  one not only to  study  nuclear processes on
Earth and in  its  atmosphere but also to  look  into stellar objects,
because  it  is  highly  penetrating  due  to  the   weakness  of  its
interactions with matter.

Progress  in  theoretical  and  experimental research has led  to  the
development  of  many  detectors  designed  to  search for and  record
terrestrial and  extraterrestrial  neutrinos. Since the energy sources
in stars  are nuclear reactions, they also emit  neutrinos. A new line
of research emerged ---  neutrino  astronomy. However, a constant flux
of low--energy neutrinos is very difficult to detect,  as evidenced by
long--term experiments to study neutrino fluxes from the Sun, the star
closest to the Earth \cite{c03,c04,c05}.

Back in  1934, Baade and  Zwicky \cite{c06} suggested the existence of
neutron stars and came  up with the idea that these are  formed during
supernova explosions.  Thirty  years  later,  in  1965, Zel'dovich and
Gusseinov \cite{c07}  concluded  that  a  short  burst of high--energy
neutrinos could arise  when matter is neutronized in the gravitational
core  collapse  of a massive star.  In  the same year, Domogatsky  and
Zatsepin  \cite{c08}   pointed  out  an  experimental  possibility  of
searching  for  such  neutrino  bursts.  In  1966,  Colgate  and White
\cite{c09}  surmised  that  neutrinos  could  play  a crucial role  in
supernova explosions. Almost concurrently, the discovery of pulsars in
the Crab Nebula \cite{c10}, which is the supernova remnant observed by
Chinese astronomers  in 1054, and in Vela \cite{c11}  late in the fall
of 1968  provided evidence for the  formation of neutron  stars during
supernova  explosions.  Subsequent  observations  and  data   analysis
confirmed  this  conclusion  and  gave  an  insight into the  physical
processes that underlie these phenomena \cite{c12,c13}.

In the succeeding years, both a theory for the final stages of stellar
evolution  and  experiments  to  search  for  bursts of such  collapse
neutrinos were intensively  developed.  The explosion of the supernova
SN  1987A  on  February  23,  1987, was  a  special  milestone  in the
development of the two lines of research.

The first experiments to search for neutrino bursts from gravitational
stellar core collapses were begun in  1973 by Pennsylvania-Texas-Turin
collaboration. These were three small water  Cherenkov facilities with
target masses of about 20  t  each and with particle detection  energy
thresholds of 20 MeV: the first was at the Homestake  Mine  at a depth
of 4400 meters of water equivalent (m.w.e.) \cite{c14}, the second was
at a mine in Ohayo at a depth of 1800 m.w.e. \cite{c15}, and the third
was in the road tunnel under Mont Blanc at  a  depth  of  4270  m.w.e.
\cite{c16}.  They  did not work long,  only  a few years. Although  no
expected burst of collapse neutrinos was detected, they made the first
step on the long way to 1987.

The  Baksan  underground  scintillation telescope (Northern  Caucasus,
Russia)\cite{c17}, the  LSD  scintillation  facility  under Mont Blanc
(Italy) \cite{c18}, the scintillation detector at the salt mine in the
town  of  Artemovsk (Ukraine)  \cite{c18},  and  the  water  Cherenkov
detectors  Kamiokande  II (Japan) \cite{c19} and IMB (USA)  \cite{c20}
belong to the  second  generation of collapse--neutrino detectors with
target masses of several hundred tons, which began  their operation in
1977---1980. The last  two detectors were specially designed to search
for  proton  decays, but  they  proved to  be  incapable of  recording
collapse neutrinos.

Before 1987, the theoretical  models  for the late evolutionary stages
of massive  stars  were  one--dimensional  calculations of spherically
symmetric   nonmagnetic   nonrotating    configurations.   The   basic
neutrino--radiation parameters expected during the gravitational  core
collapse of a massive  star and the cooling of a newborn  neutron star
were formulated: the  total neutrino--radiation energy is $(2 \div 5)\
10^{53}$erg, íà£, the  mean electron--neutrino energy is $(8 \div 12)$
MeV, the fraction of electron antineutrinos in the  total radiation is
$(0.16  \div  0.25)$  and  the  burst duration  is  $(10  \div  20)$ s
\cite{c21,c22,c23,c24}.

The recording of a neutrino signal from a  type--II supernova exploded
in  the  Large  Magellanic  Cloud,  a  neighboring  galaxy,  by  three
facilities, Kamiokande II  \cite{c25},  IMB \cite{c26}, and the Baksan
telescope \cite{c27}, on  February 23, 1987, first confirmed the basic
ideas of the general theoretical  pattern  of  gravitational  collapse
with the  formation of  a neutron star. At the  same time, however, it
raised many questions both for theory and for experiment.

Observations of the expanding  envelope  of the supernova remnant show
that  the  explosion was  asymmetric,  with an  unusual  shape of  the
remnant,  with  matter  mixing  \cite{c28,c29},   with   an   as   yet
undiscovered pulsar  \cite{c30} or with  a candidate for a pulsar with
unusual properties \cite{c31,c32}. All  of  this led to rapid progress
in  the  theoretical   modeling  of  processes  inside  stars  and  in
understanding the  many processes in  core collapse and in the stellar
envelope,  as  well as  to  a  complication  of  the  collapse pattern
\cite{c33,c34,c35},  particularly  when  rotation and strong  magnetic
fields are included in  the  calculations \cite{c36}. However, not all
of  the  questions have  been  answered  as  yet.  In  particular, the
stellar--envelope  ejection  mechanism  and,  what  is  important  for
experimenters,  the   complete  temporal  structure  of  the  neutrino
luminosity  are  still   problems  to  be  solved.  In  addition,  the
general--relativity \cite{c37} and weak--magnetism \cite{c38} effects,
which have not  yet  been included in the  calculations  and which may
lead to a higher neutrino  luminosity,  as well as a possible  nonzero
neutrino  mass,  which  would  cause  the  neutrino spectra to  change
\cite{c39}, are being discussed.

The development of  methods for analyzing experimental data with small
statistics and  using them to study  the neutrino signals  recorded on
February 23,  1987, by three facilities  also led authors  to conclude
that  the  neutrino radiation has a more  complex  temporal  structure
\cite{c40}.

An important lesson drawn from the history of experimental observation
of the neutrino signal from SN 1987A is that the facilities capable of
recording collapse  neutrinos  improved  and  adjusted  their physical
parameters and are combined into  a  single  network, SNEWS (SuperNova
Early Warning System), to form a coincidence trigger for their signals
(when they  appear) that will notify  observatories of other  types of
radiation of the beginning of supernova observations \cite{c41}.

In addition, it became clear that detectors with an even larger target
mass sensitive  to all types of neutrino are  required to reliably and
completely record the expected Galactic  neutrino  burst.  This is how
the third--generation  neutrino detectors with a characteristic target
mass of about  1 kt  or higher  and  with low  event detection  energy
thresholds (about  5 MeV) emerged. When  a Galactic burst  of collapse
neutrinos occurs, the most powerful of  these  detectors  will  record
over 8500 neutrinos of all types (SuperKamiokande, the  target mass is
32 kt of water)  \cite{c42}, about 1000 neutrinos of all types  in the
SNO detector with heavy and light  water (the total target mass is 2.4
kt)  \cite{c43}, and over  350  neutrinos  of  all types  in  the  LVD
scintillation detector  \cite{c44}. Apart from the long--operating LSD
\cite{c18} and  MACRO  \cite{c45} scintillation facilities, which will
also record about  100  or more  Galactic  neutrinos each, the  Amanda
facility in the  ice  at the South Pole  in  Antarctica \cite{c46} was
included in the network of collapse-- neutrino observations in 1997. A
number of new detectors are also  prepared to be launched or are being
designed \cite{c47}.

Thus, of  particular importance is the  so unexpected phenomenon  of a
type--II or, possibly, type--Ib/c supernova \cite{c48}  in our Galaxy.
At  least  seven  neutrino  detectors,  one  of which  is  the  Baksan
underground scintillation telescope, were prepared for its recording.

\section{The Baksan Underground Scintillation Telescope}

The Baksan underground  scintillation telescope is an instrument for a
broad range of scientific research, with one of  its experiments being
the search for neutrino burst from supernova explosions. The telescope
is located in the Northern Caucasus,  at the foot of the Mount Elbrus,
in the tunnel under the Andyrchi  Mountain at a depth of 850 m.w.e.. A
general view of the telescope is schematically shown in Fig.1, and the
facility is described in detail in \cite{c49}. Here, we briefly recall
its basic parameters.

\begin{figure}
\epsfysize=0.25\vsize
\centerline{\epsfbox{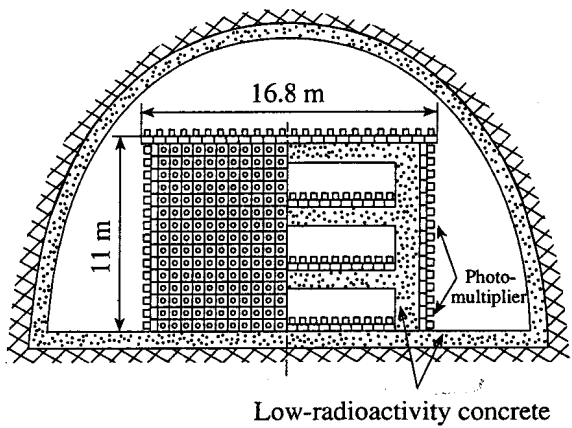}}
\caption{A schematic  view  of  the  Baksan  underground scintillation
telescope.}
\end{figure}

The telescope consists of 3150  standard  detectors.  These  detectors
form a closed configuration with two  internal  layers  that  comprise
four  vertical  planes  and  four  horizontal  planes.  Each  standard
detector is $70 \times 70 \times  30$ á¬${}^3$ in size, is filled with
an organic $C_n H_{2n+2}$, ($n \sim 9$) scintillator, and is viewed by
one photomultiplier  with a photocathode  diameter of 15 cm. The total
target  mass is  330 t,  and  the target  mass enclosed  in the  three
internal (starting  from the lower horizontal  plane) layers is  130 t
(1200  standard  detectors). The  charged--particle  detection  energy
thresholds are 8 and 10  MeV  for the horizontal and vertical  planes,
respectively. The energy reference for standard--detector measurements
is the amplitude of the most probable energy  release when cosmic--ray
muons pass through it  (50  MeV). Accordingly, the detector thresholds
are given in fractions of the amplitude of this energy release.

An  intense  burst  of neutrinos of  all  types  is  expected during a
type--II supernova explosion. However, the vast majority of the events
recorded  with  the  Baksan  telescope  will  be  produced  in inverse
beta--decay reactions,
\begin{eqnarray*}
\bar{\nu_e} + p \rightarrow n + e^{+}.
\end{eqnarray*}
The contribution  from the remaining types  of neutrinos to  the total
recorded signal will be minor \cite{c49}.

Since the telescope is located at a relatively small depth,  a special
event--selection method is used to significantly reduce the background
from the cosmic--ray muons that cross the facility  in our experiment.
This methods is  based on an  enormous energy difference  between  the
muons and the  positrons  formed from collapse electron antineutrinos.
Passing through  the telescope, the muons  leave a track  from several
triggered  detectors,  while the  positrons  formed  in  a  particular
detector  will  lose all  of  their  energy  almost  entirely  in this
detector.  Thus,  the main event--selection criterion in our  program,
--- one and only one detector from 3150, ---- implies the selection of
single standard--detector  triggerings;  i.e.,  at the program marker,
data on  the state of all devices  in the  facility at the  triggering
time  of  one  and  only  one detector in it is written to the  online
computer memory and then to storage.

As a result,  the  neutrino signal from a  supernova  explosion on the
facility  will appear  as  a series of  events  from singly  triggered
detectors during the neutrino burst.

The count rate from  single  background events, the statistical clumps
of which  can imitate the expected signal, was  not constant with time
at  the  beginning   of  the  experiment  on  the  telescope,  because
operations  were  continuously performed  to  remove  one  or  another
background source, which caused a  continuous  reduction  in the total
count rate of  single  events. Figure 2 shows  time  variations in the
yearly  mean  count rate of single detector  triggerings  summed  over
three internal planes (the lower and  the next two planes in Fig.1) in
which 130 t of scintillator are contained in 1200 detectors.  The data
from this part  of  the  facility, i.e., the total  number  of  single
triggerings  within  a given time interval,  act  as a trigger in  the
offline analysis of information from  the  telescope:  when an unusual
signal  was  detected in this part  of  the telescope (i.e., when  the
number of events was larger than a given number), information from the
entire facility was analyzed offline.

\begin{figure}
\epsfysize=0.25\vsize
\centerline{\epsfbox{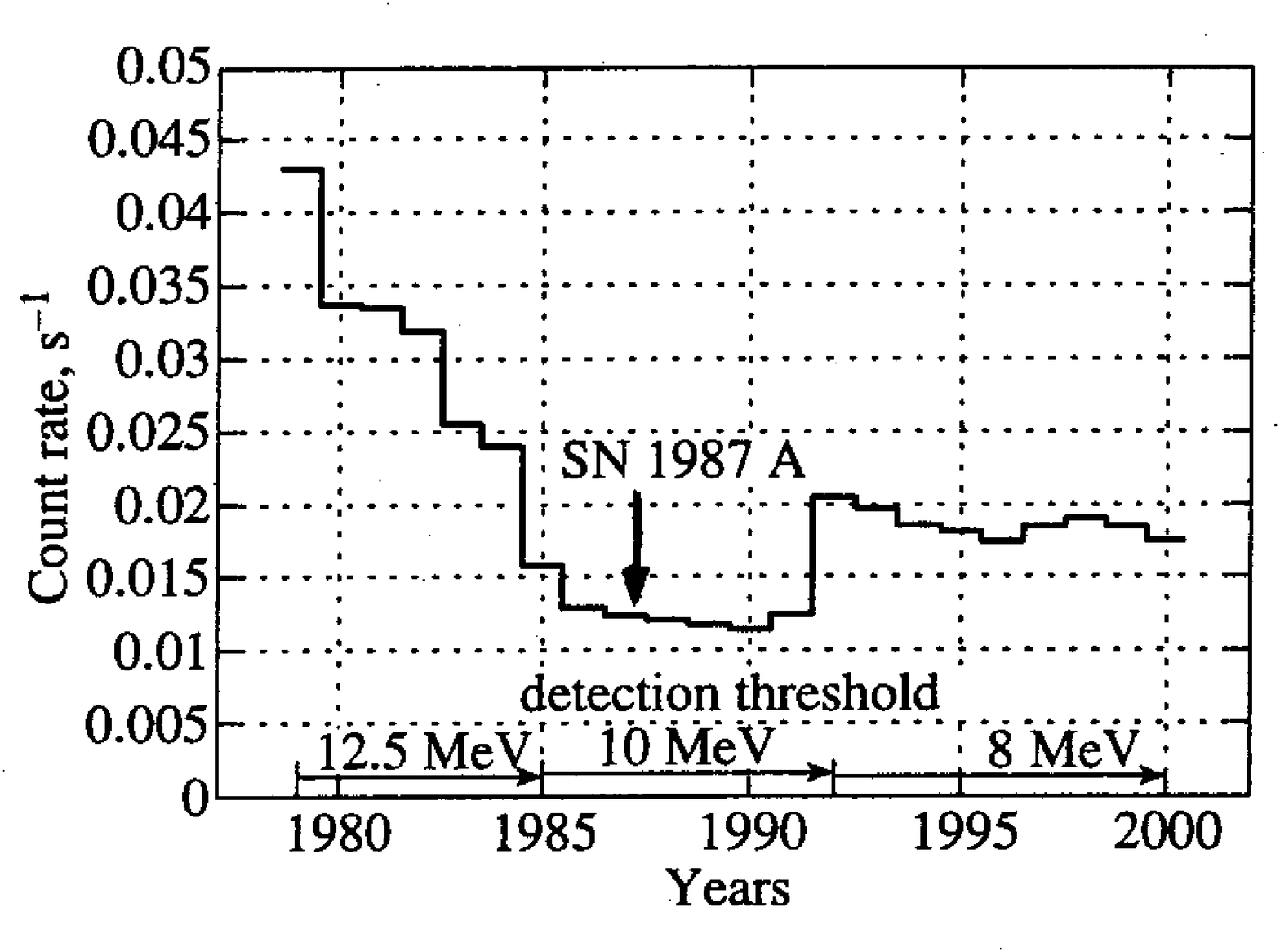}}
\caption{Time  variations  in  the  yearly mean count rate  of  single
triggerings of 1200 detectors  in  the three telescope internal planes
(the target mass is 130t). The particle detection energy thresholds in
different years are  shown. The arrow  indicates the explosion  of  SN
1987A on February 23, 1987.}
\end{figure}

Figure 2 also indicates the time intervals when the particle detection
energy threshold on  the  telescope was reduced. As  we  see from this
figure, the background after 1985, when the main  operations to remove
the strongest noise were finished, was  virtually  constant  within  a
certain detection  energy  threshold.  Its subsequent modest reduction
resulted  from  the continuous  operations  to  improve  the  physical
parameters of the facility.

The time when the explosion of  SN 1987A occurred is shown in the same
figure.

If the core of a massive star collapses at the Galactic center to form
a neutron star, then one might  expect over 50 events on the telescope
from  the  interactions  of  electron  antineutrinos   in  the  130--t
scintillator or  over 100 events  in the entire telescope. The current
information is  analyzed by the  method of a 20--s--long time interval
sliding from event to event with the total threshold number  of events
in the  signal  equal  to 5. For detection of a signal with a given or
larger number of  events,  40  rather than 20 s  of  information  with
complete  data  on  each  event (coordinates on the  telescope,  time,
energy amplitude, the number of detector  triggerings  in  the  signal
over  the   current   day,  and  photomultiplier--pulse  duration)  is
analyzed. A  signal would be considered to be  a serious candidate for
collapse detection  if nine  or more events were observed  for 20 s in
the telescope  target mass of 130 t (internal  layers). No such signal
was  detected  over   the  entire  period  of  observations  with  the
telescope.

The  telescope  was constructed late in 1977.  However,  the  physical
experiments began later, because it took time for us to figure out how
the  instrument  behaved.  The  facility  has  been  operating  almost
continuously under the program  of  search for collapse neutrinos (the
program was called  Collapse) since the mid--1980. One maintenance day
per week, when the necessary repair  and  maintenance  operations  are
performed on the facility, and  the  emergency  situations of seasonal
power cuts or online computer replacement constitute an exception. So,
the  total  time of  Galactic  observation accounts  for  90\% of  the
calender time.

Thus, information from the telescope was accumulated under the program
of search  for collapse neutrinos,  but there was no expected neutrino
signal. The history of detecting the neutrino signal from SN  1987A in
the Baksan data and the related dramatic searches for the cause of the
then unexpected clock error have long been published \cite{c28,c49}. A
new clock  with a self--contained power  supply has been  in operation
since February 1988, providing a 1--ms  accuracy  of  determining  the
absolute time.

\section{Results of the Observations}

When the Baksan experiment to search for collapse neutrinos began, the
frequency of expected signals was of great importance. Observations of
historical supernovae in our Galaxy,  supernovae  in  other  galaxies,
stellar statistics, pulsar statistics, and even the thermoluminescence
of  samples  of  bottom  sediments  were  used in  the  literature  to
determine  the  time interval  between  events  in  which  supernovae,
neutron stars, or black holes were expected to be formed.

At  that  time, the  most  optimistic estimate  of  this interval  was
obtained from an analysis based on observations of  149 pulsars, which
predicted  the  birth  of  one  pulsar  approximately every six  years
\cite{c50}.  Although  a  later  reanalysis of the pulsar  birth  rate
yielded  a  different  estimate  of  this  interval,  about  30  years
\cite{c51}, we  still hoped for a positive result  at the beginning of
our experiment, especially since the  last  supernova  was observed on
Earth 400 years ago.

Another  optimistic   result   was   obtained   by  analyzing  stellar
statistics:  when  calculating  the  total  star  death rate with  the
ultimate formation of  pulsars, black holes, supernovae, and any other
possibilities,  except  white dwarf, the interval between such  events
was found to be $9^{+2}_{-3}$ years \cite{c52}. The  estimates for the
rate  of  type--II supernovae in our  Galaxy  by other authors give  a
large spread in the predictions of  its mean value. This appears to be
attributable to a large number of assumptions used and approximations,
for  example,  about  the  luminosity  of  the Galaxy \cite{c53},  its
morphological type  \cite{c54},  the initial mass function \cite{c55},
the  fraction  of   detected   pulsars  \cite{c56},  and  many  others
\cite{c57,c58,c59,c60}.

Subsequently,  the  relationships  between  supernova  statistics  and
galactic  evolution  were  analyzed  and  it  was  concluded  that the
supernova  rates  were  functions  of cosmic time  \cite{c61,c62}.  In
addition, a study of the light curves for SNe Ib/c  showed  them to be
similar to the light curves  for  type--II  supernovae \cite{c48}, led
one to  conclude that  the SN Ib/c progenitor was  a massive star, and
suggested  that  neutron  stars  could  also  be  formed  during their
explosions.

Recent estimates of the supernova  rate  from  stellar statistics show
that the following number of such events may be expected in our Galaxy
\cite{c63,c64}: $2^{+1}_{-1}$ for SNe Ib/c and  $12^{+6}_{-6}$ for SNe
II in  1000 years. Recently,  however, the same authors improved these
Galactic values \cite{c65}: $1.5^{+1.0}_{-1.0}$ for SNe II $+$ Ib/c in
100 years or about one supernova  with the formation of a neutron star
every  50  years. The latter value  is  almost equal to this  interval
previously estimated to be 10---50 years  \cite{c66}  when  the  above
possible sources of  discrepancies are taken into account. In general,
it  is  clear   that  the  sought--for  rate  estimated  from  stellar
statistics lies within  the range $10^{-2}  - 3 \cdot  10^{-2}$  per
year or with a mean estimate of the interval between  supernovae equal
to $47^{+12}_{-12}$ years \cite{c67}.

In recent years, it has been found from the statistics of pulsars that
they  are  formed, on average, once every  60--330  years  \cite{c68},
although  it  is  still  unclear  in what  systems  the  pulsars  of a
particular class are born \cite{c69,c70}.

Thus,  recent  estimates  for  the rates of supernova  explosions  and
pulsar formation lead us to conclude  that this event is rare and that
the spread  in estimates is large. Therefore, it  becomes all the more
necessary  to  obtain  the  result from direct observations  of  these
events with neutrino detectors.

The Baksan telescope  has  been observing  the  Galaxy since June  30,
1980.  Because  of  all  the  operations  performed  to  increase  its
sensitivity to the expected burst of collapse neutrinos, the telescope
views the entire Galaxy \cite{c49}. The calendar time  of the Collapse
experiment is 19.75 years, while the total live observing time is 17.6
years. No signal, except SN 1987A in the Large Magellanic  Cloud, that
could  be  reliably  interpreted  as  a  burst  of  Galactic  electron
antineutrinos was detected with the facility over this period.

An upper bound on the mean frequency of gravitational collapses in the
Galaxy can be obtained from the  observing time. If we denote the mean
frequency of collapses by $f_{collapse}$ and if we assume that, first,
their frequency  (as rare events) obeys  the Poisson law  and, second,
the probability of missing the signal is less than  10  \%  at  90  \%
confidence, then we derive the following inequality for the total live
observing time T = 17.6 years:
\begin{eqnarray*}
e^{-f_{collapse}   T}   <  0.1,
\end{eqnarray*}
whose solution  $for f_{collapse}$ yields  a bound on the frequency of
collapses in the Galaxy,
\begin{eqnarray*}
f_{collapse} (90 \% \mbox{confidence}) < 0.13 \mbox{yr}^{-1}.
\end{eqnarray*}

Hence,  the  mean time  interval  $\Delta  T_{collapse}$  between  the
expected Galactic events exceeds
\begin{eqnarray*}
\Delta T_{collapse} (90 \% \mbox{confidence}) > 7.7 \mbox{years}.
\end{eqnarray*}

Thus,  the  first  twenty years on  the  path  to  detecting the first
neutrino burst  in our Galaxy have  been traversed, although  this may
prove to be only a small part of the required path.

\section{Acknowledgments}

We wish  to thank the staff of the  Baksan neutrino observatory, those
who  are  still  working  and  those who have left it, for a long  and
fruitful collaboration in this experiment. This study was supported by
the Russian Foundation for Basic Research (project no. 00-02-17778).

\end{document}